# Elucidating the Role of Stacking Faults in TlGaSe$_2$ on its Thermoelectric Properties

*Tigran Simonian$^{†§}$\*, Ahin Roy$^{†§ ]}$, Akash Bajaj$^{‡§}$, Rui Dong$^{‡§}$, Zheng Lei$^{Δ}$, Zdeněk Sofer$^{Δ}$, Stefano Sanvito$^{‡§}$, Valeria Nicolosi$^{†§}$\**

† School of Chemistry, Trinity College Dublin, College Green Dublin 2, D02 PN40, Ireland

‡ School of Physics, Trinity College Dublin, College Green Dublin 2, D02 PN40, Ireland

Δ Department of Inorganic Chemistry, University of Chemistry and Technology Prague, Technická 5, 166 28 Prague 6, Czech Republic

§ Center for Research on Adaptive Nanostructures and Nanodevices (CRANN), Trinity College Dublin, 43 Pearse St, Dublin 2, D02 W085, Ireland

$^{]}$Materials Science Centre, Indian Institute of Technology Kharagpur, Kharagpur, West Bengal, IN721302

\*Correspondence should be addressed to T.S. (simoniat@tcd.ie) and V.N. (nicolov@tcd.ie)

ABSTRACT

Thermoelectric materials are of great interest for heat energy harvesting applications. One such promising material is TlGaSe$_2$, a *p*-type semiconducting ternary chalcogenide. Recent reports show it can be processed as a thin film, opening the door for large-scale commercialization. However, TlGaSe$_2$ is prone to stacking faults along the [001] stacking direction and their role in its thermoelectric properties has not been understood to date. Herein, TlGaSe$_2$ is investigated via (scanning) transmission electron microscopy and first-principles calculations. Stacking faults are found to be present throughout the material, as density functional theory calculations reveal a lack of preferential stacking order. Electron transport calculations show an enhancement of thermoelectric power factors when stacking faults are present. This implies the presence of stacking faults is key to the material's excellent thermoelectric properties along the [001] stacking direction, which can be further enhanced by doping the material to hole carrier concentrations to ~$10^{19}$ cm$^{-3}$.

Thermal management is a prescient issue in electronic devices such as the high-performance CPUs and GPUs, which are the foundation of current advances in machine learning [1–3], computational chemistry [4–7], climate modelling [8–11], etc., as inefficient heat transfer can lead to overheating and damage to the device. This is especially important as the transistors in such devices approach the nanometer and atomic scales [12]. A mitigation procedure for overheating is "thermal throttling", where the clock speed of the device is reduced to protect it from potential damage due to overheating at the cost of performance [13–16]. To avoid this and retain performance, the governance of this excess heat is often managed through a combination of heat sinks, fans, or liquid-cooling systems, which require additional electrical energy leading to increased operating costs. Conversion of this excess heat back into electrical energy would provide a means of powering the cooling systems for the devices, reducing overall costs and increasing efficiency.

Thermoelectric materials could be used to achieve this goal. These materials have low thermal conductivity yet possess high electrical conductivity, allowing them to convert heat energy into electrical energy when there is a thermal gradient across the material [17–20]. The performance of thermoelectric materials is compared via the dimensionless figure of merit *ZT*, which is defined as:

$$ZT = \frac{S^2 \sigma}{\kappa} T,$$

where *S* is the Seebeck coefficient (V K$^{-1}$), $\sigma$ is the electrical conductivity [($\Omega$ m)$^{-1}$], $\kappa$ is the thermal conductivity [W (m K)$^{-1}$], and *T* is the temperature (K) [21]. For comparison, bulk Si has a *ZT* of approximately 0.01 at 300 K due to its high thermal and electrical conductivity [22]. Many commercially available thermoelectric devices are made from PbTe- and Bi$_2$Te$_3$-based alloys, with *ZT* ~ 0.7 - 0.8 at 300 K [23–27], but these devices have low conversion efficiencies of only 8% at ambient temperatures [24,28]. The family of Bi$_x$Te$_y$ nanowires has some of the highest reported *ZT* values at room temperature, where *ZT* ~ 0.8 - 1.5, but manufacturing devices with these materials at scale has been a challenge until recently [29,30].

Other materials such as clathrates [31] and skutterudites [32] follow the concept of "phonon glass, electron crystal" in order to achieve low thermal conductivity while maintaining electrical performance [25,26,33]. These materials contain weakly bound atoms in the cavities of the unit cell, enabling anharmonic "rattling" that, in turn, leads to the phonon softening [34–37]. While these materials possess *ZT* > 1.3, these typically have high operating temperatures of at least 600 K [25,26,33] making them unsuitable for many consumer-grade applications [25].

Another potential material for high-*ZT*, ambient temperature thermoelectric devices is the ternary chalcogenide TlGaSe$_2$, a *p*-type semiconductor with a *ZT* ~ 0.8 at 300 K [38–40]. TlGaSe$_2$ has a band gap of ~ 1.95 - 2.2 eV [39–44], far larger than that of the commercial PbTe- or Bi$_2$Te$_3$-based alloy thermoelectric devices [45,46], allowing for optoelectronic applications such as X-ray and gamma-ray photodetectors [42,47].

Recently, thin films of TlGaSe$_2$ have been reported [47], allowing for the production of thermoelectric devices at large scales. The layered nature of the material leads to highly anisotropic thermoelectric properties [39,48], which may be exploited for controlled thermal transfer through a device [49,50]. This layering also means that the material is susceptible to planar defects such as stacking faults, as it had been noted in previous reports [51,52]. However, it has yet to be established if these stacking faults are intrinsic to the system or are a by-product of external stimuli such as heat.

Stacking faults in other layered materials have been shown to dampen the thermal conductivity along the stacking direction, which can increase the *ZT* of the material [53–56]. While the presence of Se vacancies is theorized to greatly hamper the thermoelectric properties of TlGaSe$_2$, the role of stacking faults on this property has yet to be established to date [38]. In this work, TlGaSe$_2$ was characterized via high-angle annular dark field scanning transmission electron microscopy (HAADF-STEM) imaging and selected area electron diffraction (SAED) patterns, which were then correlated to simulations. The nature of the stacking faults in TlGaSe$_2$ was investigated using density functional theory (DFT) calculations of the stacking fault energies. The role of these faults on the thermoelectric properties was furthermore probed via electron transport calculations, highlighting an enhancement due to the presence of the stacking faults and when the carrier concentration is increased.

Bulk crystals of TlGaSe$_2$ were grown via vacuum melt growth and its monoclinic structure (space group *C12/c1*; ICSD – 17397) [57] was confirmed via X-ray diffraction (XRD) (Figure 1a-b, see experimental section in Supporting Information). The lattice parameters were determined to be $a = b = 10.772$ Å, c = 15.636 Å, β = 100.06°.

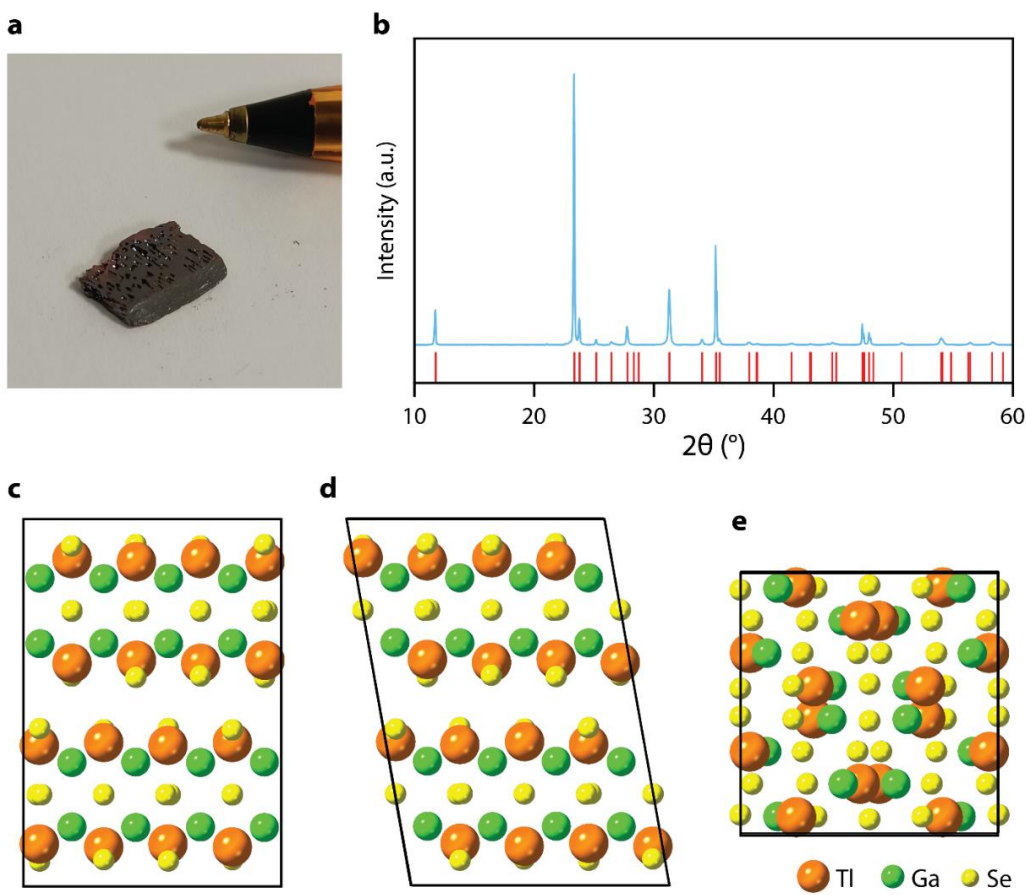

**Figure 1**: (a) Photograph of bulk TlGaSe$_2$, with tip of ballpoint pen for scale; (b) XRD pattern of TlGaSe$_2$ (blue), along with appropriate fitting of the structure (red) from ICSD – 17397; (c - e) Unit cell of TlGaSe$_2$, shown along [100], [010] and [001] zone axes, respectively.

The unit cell consists of Ga$_4$Se$_{10}$ units, each containing four corner-sharing GaSe$_2$ tetrahedra (Figure 1c - e). These form two layers within each unit cell which are aligned parallel to the

(001) plane, and perpendicular to the [114] direction. The layers are connected to each other through the $Tl^+$ ions located in the center of the $Se_6$ trigonal cavities between the layers [51,52,58,59].

HAADF-STEM imaging of $TlGaSe_2$ samples along the [1$\bar{1}$0] zone axis (Figure 2) highlights the layered nature of the material. Following along the stacking direction of [001] reveals many stacking faults throughout the sample (Figure 2c,d). Multislice simulation of the HAADF-STEM images (Figure 2b,d) confirms the presence of stacking faults. Though some short-range ordering of the layers was present, no long-range ordering was observed in any of the samples of $TlGaSe_2$ studied, which is in line with previous literature [51,52].

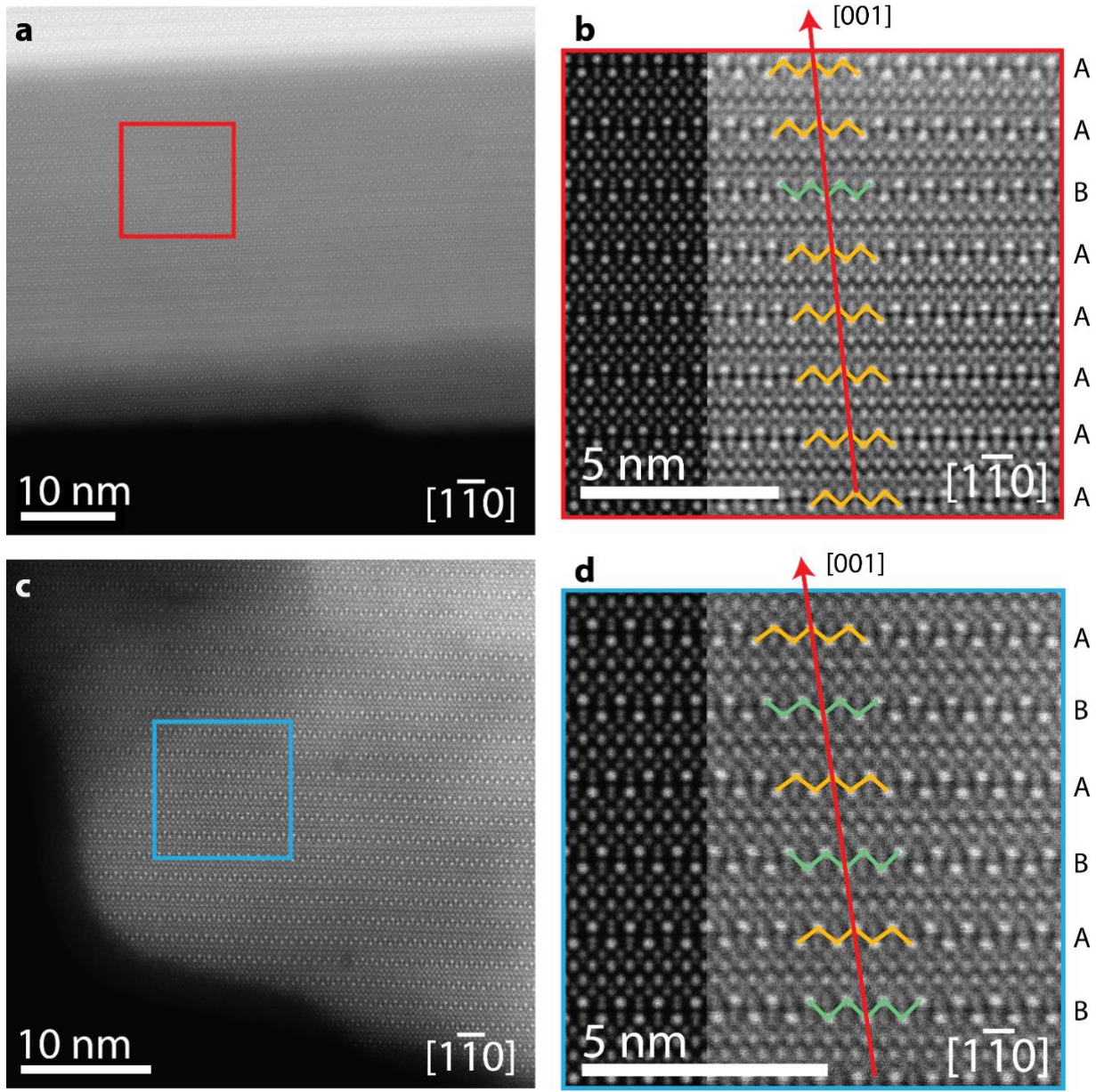

**Figure 2**: HAADF-STEM images (a,c) of TlGaSe$_2$ samples along the [1$\bar{1}$0] zone axis, with magnified regions shown in (b) and (d), respectively. Red arrows are along the [001] direction, highlighting the stacking faults along the stacking direction. The stacking order in (b) and (d) is shown along the right-hand side with green zig-zag lines to highlight the stacking faults, while multislice simulations of each image are shown along the left-hand side.

As stacking order in TlGaSe$_2$ exists between unit cells along the [001] direction and appears in one of two forms: *AA* and *AB* (Figure 3a,b). The *AA*-stacking is considered the bulk stacking mode and can be described as a pseudotranslation of a unit cell with respect to the cell below it along the [110] direction:

$$s_1 = \frac{1}{4}\boldsymbol{a} + \frac{1}{4}\boldsymbol{b},$$

where $\boldsymbol{a}$ and $\boldsymbol{b}$ are the lattice vectors. Stacking faults in TlGaSe$_2$ are classified as *AB*-stacking (Figure 3b) when there is a 90° rotation of a unit cell with respect to the cell below it about the [1$\bar{1}$4] direction. Conveniently, this too can be considered a pseudotranslation along the [110] direction:

$$s_2 = \frac{1}{4}\boldsymbol{a} - \frac{1}{4}\boldsymbol{b}$$

To corroborate the short-range order seen in the HAADF-STEM imaging, SAED was performed on the same area of the sample shown in Figure 2a. Diffuse streaking was observed along the (*hhl*) directions where $h$ = odd (Figure 3c) [51]. The presence of streaks in these diffraction patterns can indicate the presence of stacking faults in the structure [60]. As the number of stacking faults increases, the number of additional spots should increase as the diffuse intensities are split symmetrically around the positions of the Bragg intensities with $h\bar{h}l$. However, low numbers of stacking faults imply extended local domains, which can lead to the smearing or streaking of the spots along the $c^*$ direction [52,61].

The lack of fully continuous streaks in Figure 3c implies some short-range order to the stacking of the layers in this part of the sample [51]. To confirm this conjecture, a TlGaSe$_2$ model consisting of the stacking order in Figure 2a was used to simulate a SAED pattern using dynamical diffraction simulations (Figure 3d; see experimental section in Supporting Information) [51,52,61]. The simulated SAED aligns closely with the experimental SAED, however, some discrepancies in the streaking do exist. This is because the smallest SAED aperture available on the TEM used was 10 µm in diameter, an area at least 10x larger than that seen in the HAADF-STEM image in Figure 2a. The area of interest was along the edge of the sample, and the sample position and aperture were aligned such that the field of view only contained this edge. However, ensuring this was difficult and hence, parts of the sample beyond the area of interest would have been exposed to the electron beam, contributing their stacking order to the diffraction pattern. Nonetheless, the streaking does confirm the short-range nature of the stacking order of $TlGaSe_2$ as seen in Figure 2 and as had been previously noted [51,52].

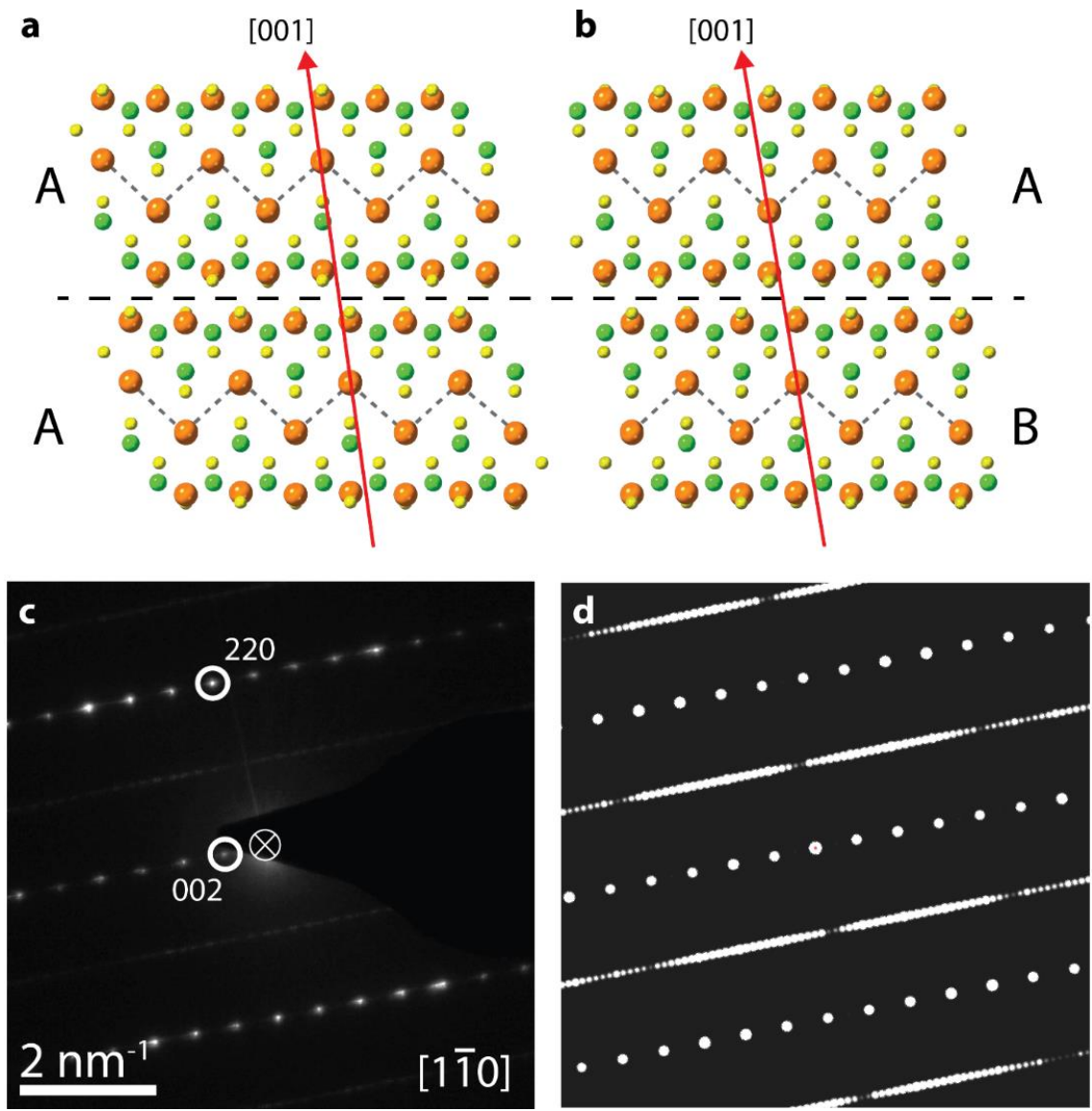

**Figure 3**: AA (a) and AB (b) stacking of TlGaSe$_2$ seen along [1$\bar{1}$0] zone axis. Red arrows showing the [001] direction and polyhedra around GaSe$_2$ units are shown to highlight the stacking order. (c) Selected area electron diffraction (SAED) pattern of TlGaSe$_2$ along [1$\bar{1}$0] zone axis. (d) Dynamical diffraction simulation of (c) along the same zone axis and same scale. The circle with a cross in (c) indicates the central beam, which has been covered by the beam blocker.

The ubiquity of these stacking faults required further examination as it was not clear whether they are intrinsic to the system (i.e. appear at the growth stage) or must be extrinsically induced via external stimuli such as heating, etc. To investigate this aspect, the stacking fault energy was calculated via DFT using the following equation:

$$\gamma_{SFE} = \frac{E_{bulk} - E_{fault}}{2A},$$

where $E_{bulk}$ is the ground state energy of the bulk system, $E_{fault}$ is the ground state energy of the faulted system, and $A$ is the area of the stacking fault [62,63]. Here, the area of the stacking fault is the area of the (001) plane of a unit cell. The ground state energy per unit cell of the bulk stacking order (*AA*) and a fully faulted (*AB*) system were calculated (see simulation section in Supporting Information), with the resulting stacking fault energies evaluated using the equation above. To establish whether an equilibrium stacking fault distance may exist, an *AAB*-ordered structure was also trailed.

The stacking fault energies are calculated to be -25.46 mJ m$^{-2}$ when comparing the bulk stacking to the *AB*-stacked structure. This is remarkable as such low stacking fault energies are only typically seen in high entropy alloys [64–68]. This would imply that the stacking faults in TlGaSe$_2$ are intrinsic to the material, however, preparing the sample for characterization may have also caused some of these faults to occur (see simulation section of Supporting Materials). Nonetheless, the negative value of the stacking fault energy and the minimal change in the value when an *AAB*-stacking order was used instead suggest that there is no equilibrium distance between the faults [64,66]. This can help to explain why long-range ordering was not observed in the experimental data, or in previous studies [51,52].

The intrinsic nature of the stacking faults may therefore play a role in the stability of the structure and hence the thermoelectric properties. However, this intrinsic nature makes it very cumbersome to isolate large enough sections of fault-free material to conduct electrical and thermal conductivity measurements. Trying to obtain a priori knowledge of which sections of the material are fault-free is also non-trivial. Hence density functional theory calculations were further used to qualitatively probe the role of the stacking faults on the thermoelectric properties of $TlGaSe_2$.

The numerator of the thermoelectric figure of merit $ZT$ can be used as a measure of thermoelectric power, which is useful as thermal transport measurements and calculations are non-trivial regardless of the material system [69–73]. To investigate this aspect, the electronic conductance for a bulk (*AA*-stacking) structure and one with a stacking fault were first calculated using first-principles ballistic transport calculations via DFT+NEGF (Figure S1, see simulation section in Supporting Information). There appears to be a negligible difference in the ballistic conductance along the [001] direction when a stacking fault is introduced. However, with Fermi level downshifts of at least ~0.5 eV below the neutrality point, the stacking fault does appear to suppress the conductance.

In order to gain a full understanding of the electronic properties relevant to thermoelectricity, the thermoelectric power factor, $S^2\sigma$, was calculated for both *AA*- and *AB*-type stacking of $TlGaSe_2$ within the semiclassical Boltzmann transport formalism under the constant relaxation time ($\tau_0$) approximation [74] (Figure 4a, see simulation section in Supporting Information). Under this approximation, only the Seebeck coefficient is independent of $\tau_0$. Thus, the power factor is reported using the scaled metric, $S^2\sigma/\tau_0$, at 300 K. For Fermi level downshifts of less than 0.2 eV below the neutrality point, the magnitude of $S^2\sigma/\tau_0$, as well as the difference between the two stacking systems along the [001] stacking direction (i.e. *zz*-components), is small. This would be the case for intrinsic TlGaSe$_2$ where the reported acceptor concentrations are only as high as $10^{15}$ cm$^{-3}$ [38,39]. However, with further Fermi level downshifts, especially around ~ 0.6 eV below the neutrality point, the *AB*-configuration exhibits a much larger $S^2\sigma/\tau_0$. This is especially possible if acceptor carrier concentrations of the order of $10^{19}$ cm$^{-3}$ can be achieved (Figure 4b). The effect of the *p*-type doping level on these power factors in the presence of stacking faults is also quite dramatic along the directions transverse to the stacking fault (Figures S2-S5). These calculations also show a clear anisotropy of the thermoelectric properties, where the *AB*-stacking enhances the thermoelectric power factors along the *zz*- and *xx*-directions, while it is suppressed along the *yy*-direction.

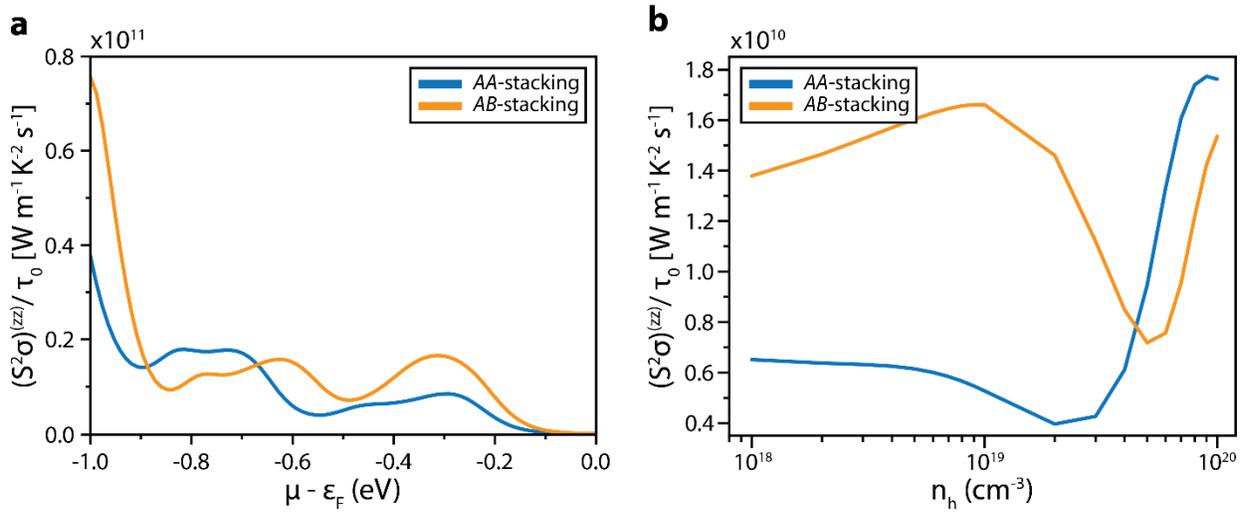

**Figure 4**: Thermoelectric power factors (a) of bulk (*AA*) (blue) and *AB*-stacking (orange) of TlGaSe$_2$ along the [001] stacking direction. The power factors as a function of hole concentration in the system is shown in (b).

In summary, the stacking faults in TlGaSe$_2$ have been found to be intrinsic to the structure, as demonstrated by their negative stacking fault energy. This implied the lack of any equilibrium stacking fault distance which was evident from the HAADF-STEM imaging and SAED studies.

The presence of stacking faults helps to explain the impressive thermoelectric properties of TlGaSe$_2$, which are then further enhanced if one were to increase the *p*-type carrier concentration via doping. It is envisioned that this research would lead to further pursuits of using TlGaSe$_2$ in a variety of thermoelectric applications.

ASSOCIATED CONTENT

**Supporting Information**.

Experimental and simulation details, electrical conductance for both with and without stacking faults using SMEAGOL, thermoelectric power factors along the *xx*- and *yy*-directions for *AA*- and *AB*-stacking, thermoelectric power factors along the *xx*- and *yy*-directions vs. hole carrier concentration for *AA*- and *AB*-stacking.

AUTHOR INFORMATION

**Corresponding Author**

Tigran Simonian - School of Chemistry, Trinity College Dublin, College Green Dublin 2, D02 PN40, Ireland; Center for Research on Adaptive Nanostructures and Nanodevices (CRANN), Trinity College Dublin, 43 Pearse St, Dublin 2, D02 W085, Ireland; ORCID: https://orcid.org/0000-0003-0259-6445 ; Email: simoniat@tcd.ie

Valeria Nicolosi - School of Chemistry, Trinity College Dublin, College Green Dublin 2, D02 PN40, Ireland; Center for Research on Adaptive Nanostructures and Nanodevices (CRANN), Trinity College Dublin, 43 Pearse St, Dublin 2, D02 W085, Ireland; ORCID: https://orcid.org/0000-0002-7637-4813 ; Email: NICOLOV@tcd.ie

**Authors**

Ahin Roy - Materials Science Centre, Indian Institute of Technology Kharagpur, Kharagpur, West Bengal, IN721302, India; ORCID: https://orcid.org/0000-0002-9515-2562

Akash Bajaj - School of Physics, Trinity College Dublin, College Green Dublin 2, D02 PN40, Ireland; Center for Research on Adaptive Nanostructures and Nanodevices (CRANN), Trinity College Dublin, 43 Pearse St, Dublin 2, D02 W085, Ireland; ORCID: https://orcid.org/0000-0002-4807-1866

Rui Dong - School of Physics, Trinity College Dublin, College Green Dublin 2, D02 PN40, Ireland; Center for Research on Adaptive Nanostructures and Nanodevices (CRANN), Trinity College Dublin, 43 Pearse St, Dublin 2, D02 W085, Ireland; ORCID: https://orcid.org/0000-0001-7008-0040

Zheng Lei - Department of Inorganic Chemistry, University of Chemistry and Technology Prague, Technická 5, 166 28 Prague 6, Czech Republic

Zdeněk Sofer - Department of Inorganic Chemistry, University of Chemistry and Technology Prague, Technická 5, 166 28 Prague 6, Czech Republic; ORCID: https://orcid.org/0000-0002-1391-4448

Stefano Sanvito - School of Physics, Trinity College Dublin, College Green Dublin 2, D02 PN40, Ireland; Center for Research on Adaptive Nanostructures and Nanodevices (CRANN), Trinity College Dublin, 43 Pearse St, Dublin 2, D02 W085, Ireland; ORCID: https://orcid.org/0000-0002-0291-715X

**Author Contributions**

T.S., A.R. and V.N. conceived the project. Z.L. and Z.S. synthesized TlGaSe$_2$ and performed XRD on the sample. T.S. and A.R. performed STEM imaging and diffraction experiments. T.S. performed the data analysis, diffraction & multislice simulations, DFT calculations on the stacking fault energies, and prepared the manuscript. All other calculations were performed by A.B. and R.D. The final manuscript was written through contributions of all authors. All authors have given approval to the final version of the manuscript.

**Notes**

The authors declare no competing financial interest.

ACKNOWLEDGMENT

T.S. and V.N. acknowledge support from the SFI Centre for Doctoral Training in Advanced Characterization of Materials (CDT-ACM) (SFI Award reference 18/EPSRC-CDT/3581). VN wishes to thank the support of the Science Foundation Ireland funded AMBER research center (Grant No. 12/RC/2278_P2), and the Frontiers for the Future award (Grant No. 20/FFP-A/8950). A.B. and S.S. acknowledge the support of Science Foundation Ireland (19/EPSRC/3605) and of the Engineering and Physical Sciences Research Council (EP/S030263/1). R.D. acknowledges the financial support of the European Innovation Council (SSlip project – grant agreement 101046693). TEM characterization was performed at the Advanced Microscopy Lab at the Centre for Research on Adaptive Nanostructures and Nanodevices. T.S. would like to acknowledge the SFI/HEA Irish Centre for High-End Computing (ICHEC) and Trinity Centre for High Performance Computing (TCHPC) for the provision of computational resources.

ABBREVIATIONS

STEM, scanning transmission electron microscopy; DFT, density functional theory; SAED, selected area electron diffraction.

(74) Madsen, G. K. H.; Carrete, J.; Verstraete, M. J. BoltzTraP2, a Program for Interpolating Band Structures and Calculating Semi-Classical Transport Coefficients. *Comput. Phys. Commun.* **2018**, *231*, 140–145. https://doi.org/10.1016/J.CPC.2018.05.010.29